\documentclass[aps,prl,reprint,superscriptaddress,
               amsmath,amssymb,amsfonts]{revtex4-2}

\usepackage[varg]{txfonts}
\usepackage{microtype}
\usepackage{bm}

\usepackage{graphicx,xcolor}
\usepackage[nolist]{acronym}

\usepackage{braket}

\definecolor{darkerblue}{RGB}{0,0,192}
\usepackage[colorlinks,
            linkcolor=darkerblue,    
            citecolor=darkerblue,  
            urlcolor=darkerblue]{hyperref}

\renewcommand{\vec}[1]{\boldsymbol{#1}}
\newcommand{\vhat}[1]{\vec{\hat{#1}}}
\newcommand{\mat}[1]{\boldsymbol{#1}}

\newcommand{\cc}[1]{{#1}^*}

\newcommand{\subref}[2]{\ref{#1}\hyperref[#1]{#2}}

\newacro{SOC}[SOC]{spin-orbit coupling}    

\begin{document}

\title{Landau Theory of Altermagnetism}

\author{Paul~A.~McClarty}
\affiliation{Max Planck Institute for the Physics of Complex Systems, N\"{o}thnitzer Str. 38, 01187 Dresden, Germany}
\email{pmcclarty@pks.mpg.de}
\author{Jeffrey~G.~Rau}
\affiliation{Department of Physics, University of Windsor, 401 Sunset Avenue, Windsor, Ontario, N9B 3P4, Canada}
\email{jrau@uwindsor.ca}

%%%%%%%%%%%%%%%%%%%%%%%%%%%%%%%%%%%%%%%%%%%%%%%%%%%%%%%%%%%%%%%%%%%%%%

\begin{abstract}
We formulate a Landau theory for altermagnets, a class of colinear compensated magnets with spin-split bands. Starting from the non-relativistic limit, this Landau theory goes beyond a conventional analysis by including spin-space symmetries, providing a simple framework for understanding the key features of this family of materials.
We find a set of multipolar secondary order parameters connecting existing ideas about the spin symmetries of these systems, their order parameters and the effect of non-zero spin-orbit coupling. 
We account for several features of canonical altermagnets such as RuO$_2$, MnTe and CuF$_2$ that go beyond symmetry alone, relating the order parameter to key observables such as magnetization, anomalous Hall conductivity and magneto-elastic and magneto-optical probes.
Finally, we comment on generalizations of our framework to a wider family of exotic magnetic systems deriving from the zero spin-orbit coupled limit.
\end{abstract}

\maketitle

%%%%%%%%%%%%%%%%%%%%%%%%%%%%%%%%%%%%%%%%%%%%%%%%%%%%%%%%%%%%%%%%%%%%%%

{\it Introduction} -- Magnetism has long been a source of novel phases and phenomena of both fundamental and technological interest. 
Many thousands of magnetic materials are known with a wide variety structures including simple colinear ferromagnets, ferrimagnets and antiferromagnets as well as more complex arrangements characterized by multiple incommensurate wavevectors~\cite{blundell2001magnetism}. 

The importance of spin-orbit coupling in magnetism is widely appreciated, through exotic transport phenomena such as the anomalous and spin Hall effects~\cite{nagaosa2010,nagaosa2010,sinova2015} as well as new physics arising from the interplay of topology and magnetism such as skyrmion physics~\cite{Fert_2017,tokura2021}, non-trivial magnon band topology~\cite{mcclarty2022} or Berry phases induced by spin chirality in the electronic band structures of itinerant magnets~\cite{nagaosa2010}. 
However, the zero spin-orbit coupled limit still holds surprises. 

One phenomenon in this setting that has captured the attention of a broad cross-section of the community~\cite{noda2016,smejkal2019,Ahn2019,naka2019,hayami2019,Hayami2020,Yuan2020,Yuan2021,Egorov2021,smejkalahe2022,reichlová2021macroscopic,Mazin_2021,Gonzalez2021,Naka2021,Shao2021,Ma2021,mazin_editorial,Jungwirth2022,smejkal2022,smejkal2022b,feng2022,bose2022,karube2022,bai2022,bhowal2022magnetic,mazin_sc,liuinverse,wangmeng2023,Mazin_2023,fedchenko2023,hariki2023,maier2023,guo2023,papaj2023,beenakker2023,zhanghuneupert2023,steward2023,zhu2023} is \emph{altermagnetism}. Following the unexpected discovery of a $d$-wave spin splitting of the Fermi surface in RuO$_2$ based on {\it ab initio} calculations~\cite{smejkal2019}, it was realized that this is one instance of a large new class of magnets defined by spin symmetries~\cite{smejkal2022}.
This spin-split band structure combines aspects of simple metallic ferromagnets and antiferromagnets with its core features borne out by experiment~\cite{bai2022,karube2022,bose2022,feng2022}.
Although \ac{SOC} is not negligible in this material, the altermagnetic spin splitting arising in the zero \ac{SOC} limit greatly exceeds any \ac{SOC} induced band gaps.
Despite having zero net moment, these bands can support spin currents with polarization depending on the orientation of the applied voltage. 
Further, an anomalous Hall response has been measured in altermagnetic materials such as RuO$_2$~\cite{feng2022} and MnTe~\cite{gonzalez2023}. 
While research into these magnets is at an early stage, there is hope that they may complete the program of antiferromagnetic spintronics~\cite{zutic2004,Jungwirth2016,Baltz2018}: realizing THz switching devices with no stray fields and with low damping spin currents. 

Despite their significant potential value in applications, there remain fundamental questions in situating these new phases of matter within the broader context of magnetism. 
From a practical standpoint, one can characterize most of the altermagnetic properties as originating from band structures with an anisotropic pattern of spin splitting in momentum space due to time-reversal symmetry breaking~\cite{smejkal2022,smejkal2022b}. 
This is in contrast to simple Stoner ferromagnets with double sheeted Fermi surfaces for the different populations of up and down spins and those of simple antiferromagnets where the Fermi surfaces are perfectly spin compensated~\cite{fazekas1999lecture,blundell2001magnetism}, as well as from frustrated isotropic antiferromagnets which can have complicated Fermi surfaces with electron and hole pockets albeit with equal spin populations~\cite{Chern2010,nagaosa2010}. 
While appealing, this phenomenology does not delineate which properties of altermagnets are robust to small symmetry allowed perturbations, and which may depend on material specific details. 

In this paper, we argue that Landau theory adapted to the zero \ac{SOC} limit captures the unique features of altermagnets. 
Starting from the definition of \citet{smejkal2022}, this Landau theory links spin symmetries to altermagnetic phenomenology including their band structures, thermodynamics and response functions, and reveals a deep connection to multipolar secondary order parameters~\cite{bhowal2022magnetic}. 
The symmetries of these multipoles relate directly to the symmetries of the spin-split bands, with the anisotropy of the electronic kinetic terms manifesting the same quadrupolar or hexadecapolar spatial structure found in the secondary order parameters, reminiscent of electronic nematic or spin-nematic phases~\cite{fradkin2010}. 
In addition, this Landau theory allows one to systematically address the effects of switching on \ac{SOC}, identifying the leading coupling to the primary order parameter and how they relate to any multipolar secondary  order parameters. 
As many of the features of altermagnets, such as the anomalous Hall conductivity \emph{only} appear when \ac{SOC} is non-zero, by approaching from this limit, we can analyze in detail how the phenomenology of altermagnets is distinguished from generic spin-orbit coupled magnets. The zero \ac{SOC} limit thus acts as  ``parent'' phase from which many of their principal features -- features that are obscured within the standard symmetry analysis -- can be understood in real materials.

%%%%%%%%%%%%%%%%%%%%%%%%%%%%%%%%%%%%%%%%%%%%%%%%%%%%%%%%%%%%%%%%%%%%%%

{\emph{Landau Theory}} -- We adopt the essential definition
put forth in \citet{smejkal2022}:
an ``ideal'' altermagnet is a spin-orbit free magnet with colinear antiferromagnetic order where the two sublattices are symmetry related by something other than translation or inversion symmetry. 
Since without \ac{SOC} spatial and spin operations can act separately, we can frame this as a statement about the spatial transformation properties of the N\'eel order parameter $\vec{N}$. 
To rephrase in this new language: in an altermagnet, $\vec{N}$ transforms as an inversion even non-trivial one-dimensional irreducible representation (irrep) under the action of the crystal point group~\cite{smejkal2022,smejkal2022b}.

To be concrete, we assume that we have a system in which we can define a uniform magnetization, $\vec{M}$, and staggered magnetization, $\vec{N}$ (both inversion even).  
In the absence of \ac{SOC} the uniform magnetization transforms
as $\Gamma_1 \otimes \Gamma^{S}_{A}$ where $\Gamma_1$ is the trivial irrep of the point group and $\Gamma^S_A$ is the (axial) vector irrep of the spin rotation group. 
We assume that $\vec{N}$ instead transforms as $\Gamma_N \otimes \Gamma^S_A$ where $\Gamma_N$ is a {non-trivial} one-dimensional irrep of the point group.
The condition that $\Gamma_N \neq \Gamma_1$ encodes the assumption of altermagnetism~\cite{smejkal2022,smejkal2022b}.

An immediate consequence is that a net magnetization is not necessarily induced in the N\'eel phase. 
To see this, we consider direct linear couplings between $\vec{N}$ and $\vec{M}$ that transform as the product $(\Gamma_1 \otimes \Gamma^S_{A}) \otimes (\Gamma_N \otimes \Gamma^S_A) = \Gamma_N \otimes (\Gamma^S_{1} \oplus \Gamma^S_A \oplus \Gamma^S_Q)$ where $\Gamma^S_{1}$ and $\Gamma^S_{Q}$ are the scalar ($\ell=0$) and quadrupolar ($\ell=2$) irreps of the spin rotation group. 
Since $\Gamma_N$ is a non-trivial irrep, these couplings are forbidden in the absence of \ac{SOC}.

We now connect this to higher multipoles.~\footnote{For simplicity we will restrict to the case relevant to most candidate materials, such as the rutiles, MnTe or CuF$_2$, where the point group includes inversion} 
Going beyond $\vec{N}$ or $\vec{M}$, we can define a time-odd, inversion even octupole, transforming like an axial vector in spin space, but a quadrupole spatially.  
Tracking spin and spatial indices separately, we can define~\cite{bhowal2022magnetic}
\begin{align}
    \label{eq:octupole}
    \vec{O}_{\mu\nu} &= \int d^3r\ r_\mu r_\nu \vec{m}(\vec{r}), 
\end{align}
where $\vec{m}(\vec{r})$ is the microscopic magnetization density. 
Note that $\vec{O}_{\mu\nu}$ transforms under spin-space symmetries as $O^{\alpha}_{\mu\nu} \rightarrow \sum_{\rho\tau\beta} S_{\alpha\beta} R_{\mu\rho} R_{\nu\tau} O^{\beta}_{\rho\tau}$ where $\mat{S}$ is a rotation in spin space and $\mat{R}$ a rotation in real space.
Other multipoles can be constructed analogously. 
This octupole transforms as
$
\vec{O}_{\mu\nu} \sim \Gamma_{Q} \otimes \Gamma^S_{A},
$
where $\Gamma_{Q}$ is the (generally reducible) representation of a spatial quadrupole. 
A linear coupling between $\vec{N}$ and $\vec{O}_{\mu\nu}$ then transforms as
$$
(\Gamma_N \otimes \Gamma^S_A) \otimes
(\Gamma_Q \otimes \Gamma^S_A) = 
(\Gamma_N \otimes \Gamma_Q) \otimes (\Gamma^S_1 \oplus \Gamma^S_A \oplus \Gamma^S_Q).
$$
Thus if $\Gamma_Q$ contains $\Gamma_N$ then $\vec{N}$ and $\vec{O}_{\mu\nu}$ can couple linearly in the absence of \ac{SOC}, and the octupole will appear as a secondary order parameter in the N\'eel phase. In the language of \citet{smejkal2022}, this would define a $d$-wave altermagnet.

We expect these multipolar secondary order parameters to be \emph{generic}; for a given symmetry there should exist a high enough rank multipole
such that its spatial part contains $\Gamma_N$. 
How do these secondary order parameters relate to the altermagnetic phenomenology? 
We first consider implications for bulk thermodynamic and transport probes, but as we will see in our discussion of the rutiles, these multipoles also connect to the symmetry of the spin-split bands.

Consider whether $\vec{N}$ can couple linearly to $\vec{M}$ once \ac{SOC} is included. 
As the Landau theory now admits magnetocrystalline anisotropy, spin and spatial transformations are coupled and the spin rotation group irreps reduce to $\Gamma^S_1 \rightarrow \Gamma_1$, $\Gamma^S_A \rightarrow \Gamma_A$ and $\Gamma^S_Q \rightarrow \Gamma_Q$.
A linear coupling between $\vec{N}$ and $\vec{M}$ thus transforms as
$$
(\Gamma_1 \otimes \Gamma_A) \otimes 
(\Gamma_N \otimes \Gamma_A) = 
\Gamma_N \otimes (\Gamma_A \otimes \Gamma_A).
$$
Using that $\Gamma_A \otimes \Gamma_A = \Gamma_1 \oplus \Gamma_A \oplus \Gamma_Q$, whether this coupling is allowed is determined by whether $\Gamma_N$ appears in the decomposition of $\Gamma_A$ or $\Gamma_Q$. 
An identical condition applies for the generation of an anomalous Hall conductivity~\cite{nagaosa2010}, corresponding to a current transverse to an applied voltage in the absence of an applied magnetic field, $J_\mu = \sum_{\mu\nu} {\sigma}^{\mu\nu}_H {E}_{\nu}$, as it transforms in the same way as $\vec{M}$. We also note that the Hall conductivity and magnetic circular dichroism, transform identically under symmetry so these conclusions also carry over to this magneto-optical probe.  

We can now connect the appearance of a multipolar secondary order parameter to $d$-wave altermagnetic phenomenology: \emph{if $\vec{N}$ couples linearly to an octupole in the absence of \ac{SOC} (and thus $\Gamma_N \subset \Gamma_Q$), then it will necessarily have a linear coupling to $\vec{M}$ and ${\sigma}_H^{\mu\nu}$ in the presence of \ac{SOC}}. 
The definition of \citet{smejkal2022} does not \emph{require} inducing an octupole, but instead can involve only higher rank multipoles, corresponding to $g$-wave or $i$-wave altermagetism. 
In those cases, the generation of weak ferromagnetism or an anomalous Hall effect can still generically persist. 
It may still be generated linearly if $\Gamma_N \subset \Gamma_A$, but will necessarily appear non-linearly otherwise.

With these core ideas outlined, we apply this framework to understand a few common examples of altermagnetic systems, including rutiles such as RuO$_2$ and hexagonal MnTe. 
We will see that by adopting this phenomenological Landau theory we can clarify the role played by multipolar  secondary order parameters, and delineate different mechanisms for the generation of characteristic responses when \ac{SOC} is included.

%%%%%%%%%%%%%%%%%%%%%%%%%%%%%%%%%%%%%%%%%%%%%%%%%%%%%%%%%%%%%%%%%%%%%%

\begin{figure}[t]  
  \begin{center}
\includegraphics[width=0.95\columnwidth]{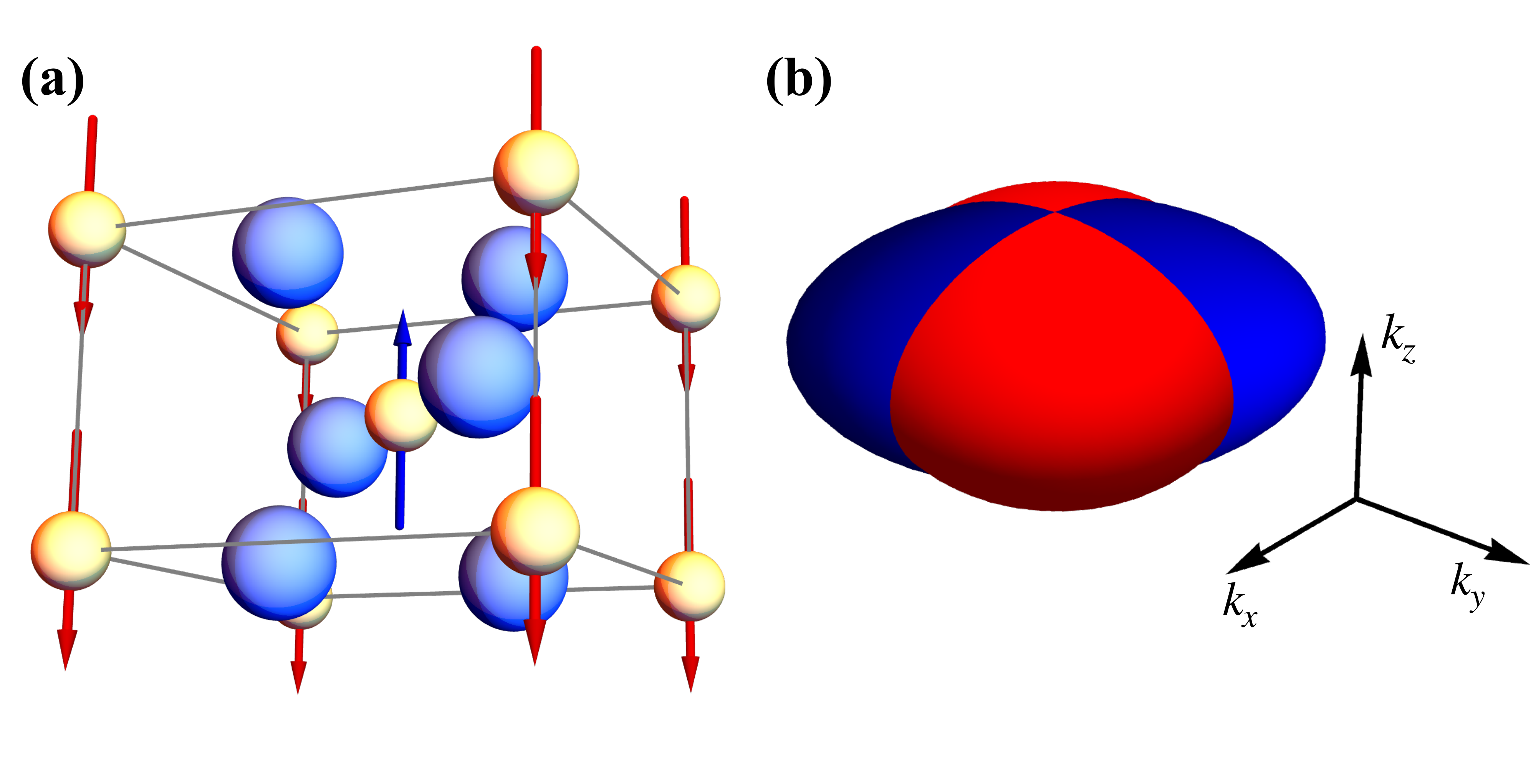}
\caption{\label{fig:rutile} 
Illustration of the (a) crystal structure of RuO$_2$ with magnetic Ru (orange) and oxygens (blue). 
(b) Fermi surface with $d$-wave spin splitting (up and down spins in blue and red, respectively) in the model of Eq.~(\ref{eq:model}).
}
  \end{center}
\end{figure}

{\it Rutile Altermagnetism} -- We begin with the canonical example of altermagnetism in rutiles with chemical formula MX$_2$ where M is the magnetic ion and X = O, F. 
The most prominent example is currently RuO$_2$ which is a metallic antiferromagnet with a simple N\'{e}el order below the magnetic ordering temperature $T_{\rm N}> 300 $K~\cite{Berlijn2017,zhu2019,Occhialini2021,bai2022,karube2022,bose2022,feng2022,lovesey2022,fedchenko2023}. 
The crystal structure belongs to tetragonal space group P4$_2/mnm$ ($\# 136$) with the Ru at Wyckoff position $2a$ and the oxygen at Wyckoff position $4f$.
The magnetic sublattice is therefore body-centred tetragonal, as shown in Fig.~\subref{fig:rutile}{(a)}. 
The space group has a generator $C_{4z}$ combined with translation through $(\tfrac{1}{2},\tfrac{1}{2},\tfrac{1}{2})$ that maps one magnetic sublattice to the other. 
The inversion centre, while present, preserves the magnetic sublattices. 
Below the magnetic ordering temperature, colinear anti-parallel moments appear on the two magnetic sublattices. 

Before delving into a phenomenological Landau description, to set the stage we consider a simple model that captures the principal features of rutile altermagnetism.
This model, introduced in Ref.~\cite{schiff2023spin}, consists of non-interacting fermions coupled to classical localized moments on the $2a$ sites through a Hund's like interaction. 
In real space the Hamiltonian is
\begin{equation}
\label{eq:model}
H = \sum_{n=1,3}\sum_a t_n^a \sum_{\langle i,j\rangle_{n,a}}c_{i\sigma}^\dagger c_{j\sigma} - J \sum_i c^\dagger_{i\alpha}\left(\boldsymbol{S}_{i}\cdot\boldsymbol{\sigma}_{\alpha\beta}\right)c_{i\beta},
\end{equation}
where $\vec{S}_i$ are the local moment directions.
One important observation is that truncating the model at nearest-neighbor $t_1$ or second-neighbor $t_2$ hoppings accidentally realizes the larger symmetry group of the underlying body-centred tetragonal lattice. 
The (lower) symmetry of the true space group $\#136$ manifests first through the presence of two inequivalent third neighbor hoppings, which generically have different amplitudes absent fine-tuning. 
The resulting band structure is such that the lowest two bands are split over most of momentum space with degeneracies along $(k,0,0)$ that arise from the spin-space symmetry of the system~\cite{smejkal2022,schiff2023spin} and with maximal splitting along $(k,k,0)$. 
As spin is a good quantum number, and the two bands correspond to electrons with polarization along $\vec{S}_i$ in spin space, the resulting splitting is the $d$-wave pattern shown in Fig.~\subref{fig:rutile}{(b)}, characteristic of a $d$-wave altermagnet~\cite{smejkal2022}.

Let us formulate an explicit Landau theory for this class of materials. 
In this system, $\vec{N}$ transforms as the non-trivial $B_{2g}$ irrep of the point group $4/mmm$ ($D_{4h}$), satisfying the definition of an altermagnet~\cite{smejkal2022,smejkal2022b}. Direct coupling between $\vec{M}$ and $\vec{N}$ is thus forbidden. More precisely the order parameter transforms under the spin point group $\boldsymbol{b}^{\infty}\otimes ^{\overline{1}}4/^{1}m^{1}m^{\overline{1}}m$~\cite{spinPointLitvin,schiff2023spin} where the superscripts refer to spin-space operations coinciding with real space generators~\footnote{Here $\boldsymbol{b}^{\infty}$ denotes the residual spin symmetry $U(1)$ around the axis of the ordered moment and also this times the operation that rotates the moments around a perpendicular axis followed by time reversal.}. 

The Landau theory for the N\'eel order parameter takes the
usual form
\begin{equation}
\Phi = a_2 \vec{N}\cdot \vec{N} + a_4 \left( \vec{N}\cdot \vec{N}  \right)^2,
\label{eq:afmlandau}
\end{equation}
enforced by spin-rotation and time-reversal symmetry. This conventional Landau theory becomes less standard when couplings to other observables are included.
For the $D_{4h}$ point group $\Gamma_Q = A_{1g}\oplus B_{1g}\oplus B_{2g}\oplus E_g$ and so the only component that transforms like $\vec{N} \sim B_{2g}$ is the $xy$ spatial quadrupole coupled with the magnetization vector. 
We thus have a linear coupling $\propto \vec{N}\cdot \vec{O}_{xy}$, as defined in Eq.~(\ref{eq:octupole}). 
It follows that $\vec{O}_{xy}$ is a secondary order parameter generated when the primary order parameter $\vec{N}$ becomes finite. 

The presence of this magnetic octupole can be directly tied to the structure of the corresponding altermagnetic band spin splitting. 
When $\vec{N} \neq 0$, hoppings and on-site terms are allowed that couple linearly to $\vec{N}$ and thus transform spatially as the same non-trivial one dimensional irrep as $\vec{N}$. As the physics is independent of spin orientation, without loss of generality, we may consider one orientation of $\vec{N}$ whereupon the spin components decouple, and the spatial dependence of the new spin-dependent terms follows the \emph{spatial} part of the multipolar secondary order parameter.
The spin-splitting of the bands thus has a form factor that mirrors the multipole induced locally. 
In the case of the rutiles, this gives a spin-splitting $\sim k_x k_y$ implying that the spin of the Fermi surface, in itinerant altermagnets, reverses in $\pi/4$ rotations about the $c$ axis, as has been established on the basis of {\it ab initio} calculations~\cite{smejkal2019}. 

The non-trivial transformation of properties of the N\'eel order parameter has implications for coupling to other observables even in the zero \ac{SOC} limit. For example, magneto-elastic couplings and piezomagnetism can be readily understood from this Landau perspective.
In the absence of \ac{SOC}, $|\vec{N}|^2$ and $|\vec{M}|^2$ couple trivially to the strains $\epsilon_{xx}+\epsilon_{yy}$ and $\epsilon_{zz}$, as dictated by the underlying tetragonal cell. 
Remaining in this non-relativistic setting, the rutile crystal exhibits non-trivial piezomagnetic couplings, even in absence of \ac{SOC}. 
To see this, note that $\vec{N}\cdot\vec{M}$ transforms like $B_{2g} \otimes \Gamma^S_1$, identical to the strain $\epsilon_{xy}$. 
In an applied field, $\vec{H}$, the Landau theory  thus admits a term of the form $\propto \epsilon_{xy}\vec{N} \cdot\vec{H}$ (see also \citet{steward2023}). 
A finite staggered magnetization in the altermagnetic phase then results in a shear distortion under an applied magnetic field. 
As noted by \citet{Dzyaloshinskii1958}, the introduction of \ac{SOC} leads to an additional coupling $\propto (\epsilon_{xz} H_y + \epsilon_{yz} H_x)N_z$.

We can relate the appearance of piezomagnetism to the underlying altermagnetism more generally. 
Considering the field transforms as $\vec{H} \sim \Gamma_1 \otimes \Gamma^S_A$ and strain as $\epsilon_{\mu\nu} \sim \Gamma_Q \otimes \Gamma^S_1$ (ignoring the uniform strain component) trilinear couplings with $\vec{N}$ transform as
$$
(\Gamma_N\otimes \Gamma^S_A)\otimes (\Gamma_1 \otimes \Gamma^S_A) \otimes (\Gamma_Q \otimes \Gamma^S_1).
$$
For the spin part we must take the $\Gamma^S_1$ component of $\Gamma^S_A \otimes \Gamma^S_A$, corresponding to $\vec{N}\cdot\vec{H}$, and then we are left with a spatial part $\Gamma_1 \otimes \Gamma_Q$. 
Thus we can conclude: \emph{if $\vec{N}$ couples linearly to an octupole in the absence of \ac{SOC}, then it will necessarily exhibit piezomagnetism in the absence of \ac{SOC}, with a trilinear coupling between $\epsilon_{\mu\nu}$, $\vec{H}$ and $\vec{N}$}, as is the case for a $d$-wave altermagnet.
Note that if an octupole is not generated, for example for $g$- or $i$-wave altermagnetism, the piezomagnetism may still be generated linearly (if $\Gamma_N \subset \Gamma_A$) or non-linearly (if $\Gamma \not\subset \Gamma_A$) as for the magnetization.

Since $\Gamma_N \subset \Gamma_Q$ here, requiring a octupolar secondary order parameter, we immediately see both weak ferromagnetism and a finite anomalous Hall response linear in $\vec{N}$ should be expected. 
More explicitly, when spin and space rotations are coupled $M_x\vhat{x}+M_y\vhat{y}$ and $N_x\vhat{x}+N_y\vhat{y}$ both transform like $E_g$ allowing a linear coupling $M_xN_y + M_y N_x$, arising microscopically from Dzyaloshinskii-Moriya exchange.
We note that a staggered magnetization along the $\vhat{z}$ direction alone does not have a linear coupling to the ferromagnetic moment. 
For the rutile, $\sigma^{xy}_H$ transforms as $A_{2g}$ and the other two components $\sigma^{yz}_H$ and $\sigma^{zx}_H$ like $E_g$. 
Thus, with \ac{SOC} we see $\sigma^{xy}_H$ only couples to $M_z$  and $\sigma^{yz}_H$, $\sigma^{zx}_H$ only to the transverse components of both the N\'{e}el vector and the magnetization. 
While the anomalous Hall effect detected in RuO$_2$ is a conventional symmetry-allowed (not fundamentally altermagnetic) response, we see that it is intimately connected to the presence of a octupolar secondary order parameter and the underlying spin group symmetries.

We have seen that the multipolar secondary order parameter in the rutile case required by $\Gamma_N \subset \Gamma_Q$ fixed many of the phenomenological altermagnetic responses expected both with and without \ac{SOC}. 
We will next consider MnTe where the quadrupole $\Gamma_Q$ does \emph{not} contain $\Gamma_N$ and the generation of higher multipoles must be considered. We also show that the magnetization, anomalous Hall conductivity as well as piezomagnetism all arise non-linearly in $\vec{N}$.

%%%%%%%%%%%%%%%%%%%%%%%%%%%%%%%%%%%%%%%%%%%%%%%%%%%%%%%%%%%%%%%%%%%%%%

\begin{figure}    
  \begin{center}
\includegraphics[width=0.95\columnwidth]{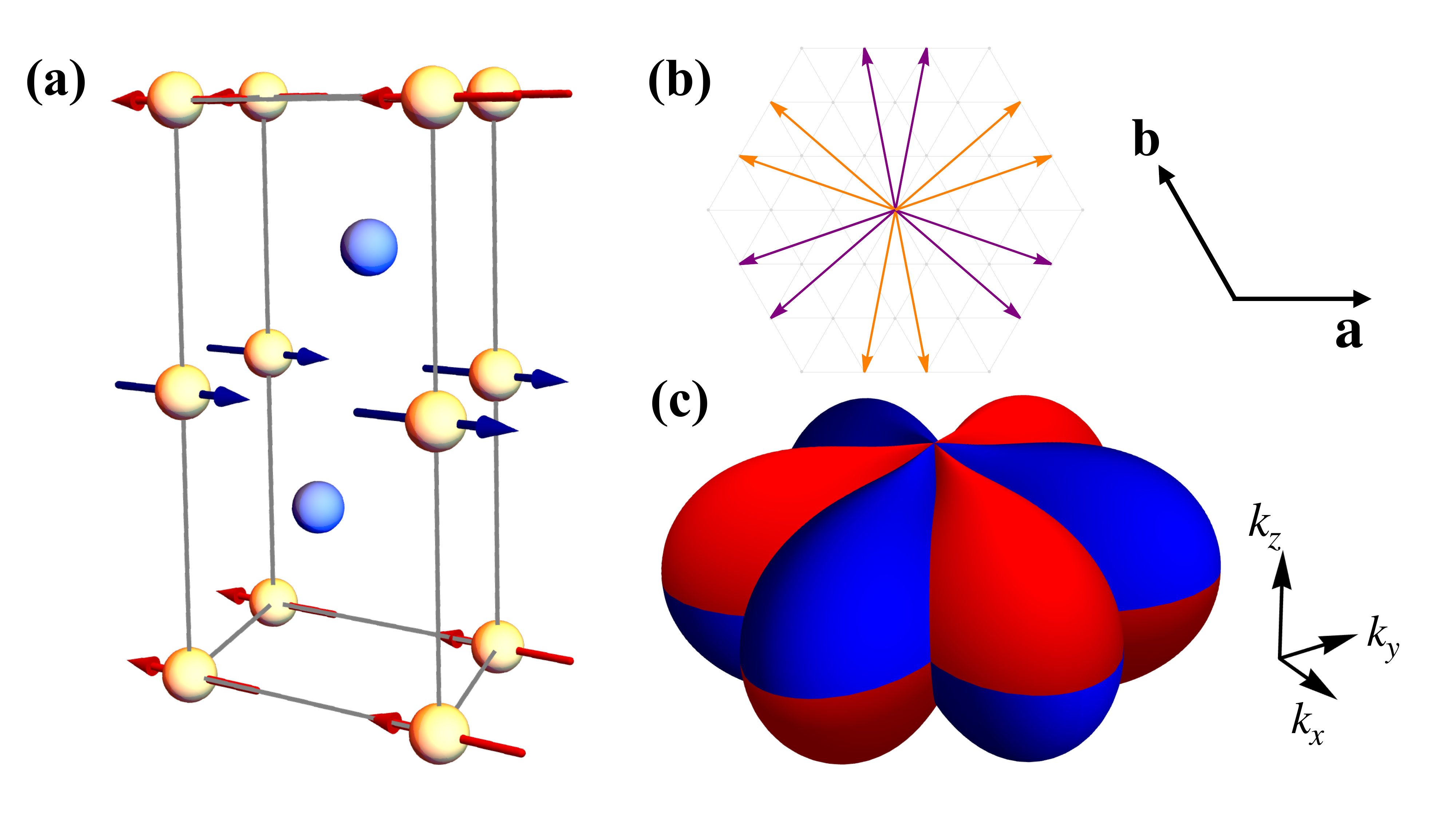}
\caption{\label{fig:mnte} 
Illustration of the key features of altermagnetic MnTe including (a) the crystal structure with magnetic Mn ions on an AA stacked triangular lattice, (b) the inequivalent bonds connecting neighboring magnetic layers along the $\vec{c}$ direction that enter into the model of Eq.~(\ref{eq:model}) and (c) and the $g$-wave spin split Fermi surface expected in weakly doped MnTe.}
  \end{center}
\end{figure}

{\it Hexagonal MnTe} -- This material~\cite{Mazin_2023,hariki2023,aoyama2023} has magnetic manganese ions on an AA stacked triangular lattice. 
The Mn ions live on the $2a$ Wyckoff positions of space group P6$_3$/$mmc$ ($\#194$) and the Te ions on the $2c$ Wyckoff positions. 
The magnetic structure is one with in-plane moments that are anti-aligned between neighboring triangular layers [see Fig.~\subref{fig:mnte}{(a)}] \cite{kunitomi1964}. 
The primary order parameter is the N\'eel vector $\vec{N}$ as in the case of the rutile altermagnet and the Landau theory is therefore identical to Eq.~(\ref{eq:afmlandau}). 
The point group is $6/mmm$ ($D_{6h}$) and $\vec{N}$ transforms as $B_{1g}$ and $\vec{M}$ as $A_{1g}$.~\footnote{The full spin point group of the order parameter is $\boldsymbol{b}^{\infty}\otimes ^{\overline{1}}6/^{\overline{1}}m^{1}m^{\overline{1}}m$} 
For MnTe, one has that $\Gamma_N \not\subset \Gamma_Q$ and thus a magnetic octupole is not induced. In the language of \citet{smejkal2022}, this is $g$-wave altermagnetism.
However, it is straightforward to see there is a higher order rank-5 magnetic multipole 
\begin{equation}
\label{eq:rank-five}
\vec{O}^4_3 \equiv \int d^3r\ (Y^4_{3}(\vec{\hat{r}})- Y^4_{-3}(\vec{\hat{r}})) \vec{m}(\vec{r}),
\end{equation}
where $Y^l_m$ is a spherical harmonic, that transforms as $B_{1g}$ -- identically to $\vec{N}$. 
A linear coupling $\propto \vec{N}\cdot  \vec{O}^4_3$ is thus allowed in the Landau theory. 
This magnetic multipole is therefore a secondary order parameter with a $g$-wave symmetry. 
The higher rank of this multipole is reflected in the nature of the band spin splitting [see Fig.~\subref{fig:mnte}{(c)}] which contains lines where the spin splitting vanishes. 
For this case a toy model can be formulated along the same lines as the rutile example but with the essential inequivalent bonds lying at relatively long range [see illustration in Fig.~\subref{fig:mnte}{(b)}]. This case highlights the potential for sufficiently long-range symmetry inequivalent hoppings to be important for altermagnetism in materials.

In contrast to the rutile case, symmetry does not permit a direct coupling between the magnetization and the staggered magnetization even in the presence of \ac{SOC} as $\Gamma_N \not\subset \Gamma_Q$ or $\Gamma_A$. 
Therefore altermagnetism does not coincide in general with weak ferromagnetism or with an anomalous Hall conductivity appearing linearly in $\vec{N}$. 
Explicit symmetry analysis reveals that coupling between $\vec{N}$ and $\vec{M}$ or ${\sigma}_H^{\mu\nu}$ appears first at \emph{third} order in $\vec{N}$. 
Restricting to an in-plane $\vec{N} = N_x\vec{\hat{x}}+N_y\vec{\hat{y}}$, as is relevant experimentally for MnTe~\cite{kunitomi1964}, one finds a single allowed coupling
\begin{equation}
\label{eq:cubic-mnte}
\sigma^{xy}_H =  a_3 N_y \left(3N_x^2-N_y^2\right) + \dots,
\end{equation}
between $\vec{N}$ and ${\sigma}_H^{\mu\nu}$ with an identical relation holding for the weak ferromagnetic moment, $M_z$. 
From the perspective of this Landau theory the generation of higher multipolar secondary order parameters thus leads to cubic (or higher) couplings between the N\'eel vector and the magnetization or Hall conductivity. 
We note that experimentally, the observed temperature dependence of the Hall signal, $\sigma^{xy}_H$, in MnTe appears convex near $T_{\rm N}$, perhaps consistent with a non-linear dependence [Eq.~(\ref{eq:cubic-mnte})] on the order parameter~\cite{gonzalez2023}.
Similarly, unlike for the rutile case, for MnTe piezomagnetism, reported in \citet{aoyama2023}, appears only in the presence of \ac{SOC} or involves non-linear couplings to $\vec{N}$ or $\vec{H}$.

%%%%%%%%%%%%%%%%%%%%%%%%%%%%%%%%%%%%%%%%%%%%%%%%%%%%%%%%%%%%%%%%%%%%%%

{\it Discussion} -- The ideas of the previous sections can be used straightforwardly to formulate Landau theories for other candidate altermagnetic materials, $d$-, $g$- and $i$-wave, with or without \ac{SOC}, as well as predict how they will couple to new physical observables.

For example, CuF$_2$ has N\'{e}el vector transforming as the $B_g$ irrep  of $C_{\rm 2h}$ ($2/m$). Since $\Gamma_Q$ contains two copies of $B_g$~\footnote{We note that here $\Gamma_N \subset \Gamma_A$ as well, providing a second route to linear couplings between $\vec{N}$ and $\vec{M}$ when \ac{SOC} is included.}, its N\'eel vector $\vec{N}$ can couple separately to the $\vec{O}_{21}$ and $\vec{O}_{21}^s$ time odd multipoles ($l=2$, $m=1$ Stevens' operators for the spatial quadrupole), with two sets of inequivalent bonds in the $xz$ and $yz$ planes. 
We can thus infer that CuF$_2$ should exhibit weak ferromagnetism and an anomalous Hall effect linear in the N\'eel order parameter as well as piezomagnetism in the absence of \ac{SOC}.

Other observables can also be treated within this framework. 
For example, one can consider the generation of spin currents~\cite{zutic2004}, characterized by a spin conductivity tensor defined through $\vec{J}^{S}_\mu = \sum_{\mu\nu} \vec{\sigma}_{S}^{\mu\nu} E_{\nu}$ where $\vec{E}$ is the electric field and the vector index encodes the spin direction. 
This transforms as $\vec{\sigma}_S^{\mu\nu} \sim (\Gamma_V \otimes \Gamma_V) \otimes \Gamma^S_A = (\Gamma_1 \oplus \Gamma_A \oplus \Gamma_Q) \otimes \Gamma^S_A$. 
Thus if $\Gamma_N \subset \Gamma_Q$ or $\Gamma_A$ then $\vec{N}$ can appear linearly in $\vec{\sigma}_S^{\mu\nu}$ in the absence of \ac{SOC}. 
For the rutile case, we would thus expect a spin conductivity $\vec{\sigma}_S^{xy} \propto \vec{N}$. 
For cases where $\Gamma_N \not\subset \Gamma_Q$ or $\Gamma_A$, like in MnTe, this would necessarily involve a higher polynomial in the N\'eel vector, $\vec{N}$.

While we have considered multipolar secondary order parameters that are even in their spatial components, when the magnetic structure lacks inversion we may find odd spatial multipoles as well. For example, point group $C_{\rm 6v}$ $(6mm)$, admits colinear antiferromagnetic spin groups, and has irreps $B_1$ and $B_2$ that allow linear couplings between certain time odd, space odd multipolar order parameters and the appropriate N\'{e}el order parameter. We leave the exploration of these multipoles for future work.

These ideas can also be generalized to non-colinear magnets. 
For example, the kagom\'e lattice with $\vec{Q}=0$, $120^\circ$ order~\cite{liumn3pt2018,hayami2020b} has a two component order parameter that can be encoded in a complex vector
$$
\vec{\Psi} = e^{+2\pi i/3 } \vec{S}_1 + e^{-2\pi i/3} \vec{S}_2 + \vec{S}_3, 
$$ 
leading to quadratic invariant $\propto \cc{\vec{\Psi}}\cdot\vec{\Psi}$. 
In the Landau theory this can couple linearly to a $d$-wave multipole in irrep $E_{2g}$ of $D_{\rm 6h}$ $(6/mmm)$ with components $k_x^2-k_y^2$ and $k_x k_y$ that itself is reflected in the spin expectation value within each band. 

%%%%%%%%%%%%%%%%%%%%%%%%%%%%%%%%%%%%%%%%%%%%%%%%%%%%%%%%%%%%%%%%%%%%%%

{\it Conclusion } -- In this paper, we have explored the application of Landau theory to altermagnets. 
This framework ties together several key ideas that have arisen in this burgeoning field including spin-split bands, spin symmetries, multipolar order parameters and the phenomenology of these materials both with and without \ac{SOC}. 
We have given examples of spin symmetric time odd multipolar order parameters that characterise these magnets as well as outlining their generalization to noncolinear altermagnetic behavior. 
These techniques are straightforwardly generalizable to the many candidate altermagnetic materials~\cite{smejkal2022,smejkal2022b} and we hope they will prove useful in sharpening predictions of altermagnetic phenomenology. 

More broadly, the considerations underpinning our Landau theory, and altermagnets viewed widely, flow from the need to generalize magnetic symmetries from the magnetic space groups to spin symmetry groups when \ac{SOC} is weak~\cite{Kitz1965,Brinkman1966,Brinkman1966b,spinGroupsLO,spinPointLitvin,corticelli2022,Liu,Liu_2022,xiao2023spin,ren2023enumeration,jiang2023enumeration,schiff2023spin,watanabe2023}. 
The induction of multipolar secondary order parameters would likely also need to be revisited in this broader context, especially as \ac{SOC} is reintroduced~\cite{hayami2018,hayami2019,Hayami2020,winkler2023}. 
Altermagnets provide a striking demonstration that there is much to be gained by thinking about novel phases, band structures and response functions in the context of these higher symmetries. 
Landau theories built from order parameters with given spin symmetries are the natural language to explore the resulting new physics and reveal how these symmetries control the altermagnetic phenomenology when \ac{SOC} is introduced.

%%%%%%%%%%%%%%%%%%%%%%%%%%%%%%%%%%%%%%%%%%%%%%%%%%%%%%%%%%%%%%%%%%%%%%

\begin{acknowledgments}
PM acknowledges useful discussions with Libor \v{S}mejkal and many interesting and informative talks at the SPICE workshop ``Altermagnetism: Emerging Opportunities in a New Magnetic Phase" 2023. 
Work at the University of Windsor (JGR) was funded by the Natural Sciences and Engineering Research Council of Canada (NSERC) (Funding Reference No. RGPIN-2020-04970).
\end{acknowledgments}

%%%%%%%%%%%%%%%%%%%%%%%%%%%%%%%%%%%%%%%%%%%%%%%%%%%%%%%%%%%%%%%%%%%%%%

\bibliography{references}

%%%%%%%%%%%%%%%%%%%%%%%%%%%%%%%%%%%%%%%%%%%%%%%%%%%%%%%%%%%%%%%%%%%%%%

\end{document}